# Combining Q&A Pair Quality and Question Relevance Features on Community-based Question Retrieval

Dong Li
School of Computer Science and Technology
Wuhan University of Technology
Wuhan, China
563408650@qq.com

Lin Li
School of Computer Science and Technology
Wuhan University of Technology
Wuhan, China
cathylilin@whut.edu.cn

*Abstract*—The Q&A community has become an important way for people to access knowledge and information from the Internet. However, the existing translation based on models does not consider the query specific semantics when assigning weights to query terms in question retrieval. So we improve the term weighting model based on the traditional topic translation model and further considering the quality characteristics of question and answer pairs, this paper proposes a community-based question retrieval method that combines question and answer on quality and question relevance (T²LM+). We have also proposed a question retrieval method based on convolutional neural networks. The results show that Compared with the relatively advanced methods, the two methods proposed in this paper increase MAP by 4.91% and 6.31%.

I. INTRODUCTION

The main research direction of Community-based Question retrieval is question relevance, which is how to measure the correlation between query and historical question and then optimize the search results. Relevance ranking is the core issue in question retrieval. It specifically refers to sort questions and answers according to their relevance with user queries. Text relevance models in the traditional information retrieval field include vector space model (VSM) [1], BM25 model [2], and language model [3].

The question retrieval model based on the statistical machine topic translation model [4-7] is a popular model for question retrieval systems currently. Clustering [8-11] is another popular method. This kind of model uses a linear combination method to integrate the language model, the translation model and the topic model into a unified framework and has achieved good results. In addition to traditional machine learning methods, deep learning based retrieval methods [12-13] have good results recent years. Cross-modal retrieval [14-15] is to find the relationship between different modal samples, and to use some modal samples to search for other modal samples of approximate semantics.

Sorting learning is a sorting method based on supervised learning, which idea is to regard the relative order relationship between documents as training data and automatically obtain the sorting model through machine learning. LambdaMART is an algorithm in sorting learning that works for many sorting scenarios and works well, companies like Bing and Facebook are using this model [16]. The current topic translation model has the same weight for the topic in the process of summing the relevance of all topics. There is no way to respond differently to lexical relevance for different query questions. For example, for a given user query "How much is a pack of white soft Chungwa cigarettes?", The top 5 search results returned by the current advanced topic translation model (T²LM) [7] are shown in Figure 1. From the table we can see that although the number of words ranked in the first candidate question is more than the others, the core word "Chungwa" in the query does not matched, which is required by the user, and so the search result is not ideal. In addition to question relevance, another recognized indicator of the impact of ranking results is the quality of questions and answers. In the case where the candidate relevance is close, the higher quality question and answer should be placed in the front position to enhance the user experience. In the research on community-based question retrieval, the quality of question-and-answer is often based not on the question-and-answer text itself, but on the use of other data in the Q&A community, such as the number of support and anti-objectives. Aiming at the above problems, we use the term weight and subject translation as the question relevance feature, and integrate the quality of question and answer pairs, and propose an optimized community-type question retrieval method. The method takes five features as input and learns training using the existing advanced model LambdaMART. The results show that compared with the relatively advanced method, the proposed method has a 4.91% increase in MAP@10.

| Rank | Candidate question |
|------|--------------------|
| 1 | 软包红牡丹 333,知道多少钱一包吗? |
| 2 | 软中华多少钱一条 2017 |
| 3 | 中华烟型号:328, 329, 330 各多少钱一包? |
| 4 | 中华多少钱一包 |
| 5 | 软中华多少钱一条 |

Fig. 1. A search example of T²LM

II. RELATED WORK

In response to the shortcomings of the word-based exact matching question relevance model, the researchers introduced the statistical machine translation model [17] into the field of information retrieval, and used the translation model to model the correlation between different words. Jeon [18] proposed a language model based on translation, the model combines the strengths of both the language model and the translation model, and introduces the translation probabilities into the language model, completing the matching between different words. Xue [19] improved the model proposed by Jeon. He integrated the maximum likelihood estimation of the question corresponding to the answer into the matching process with the query words, enriched the matching text, and proved to further the effect of question retrieval. Upgrade. Lee [20] used the more important feature words in the Q&A pair to enhance the translation model to optimize the translation probabilities between vocabularies. The training of the translation model in the above method uses the question and its corresponding answer or question and its detailed description as a monolingual parallel corpus, which makes the obtained translation model contain a large noise. In view of this, Bernhard and Gurevych [21] combined multiple high-quality monolingual parallel corpora as training data, which further improved the original retrieval model. Gao [22] integrated the interdependence of words in the question into the language model, and proposed a new language model based on word dependence. Cao [23] establishes a more complex language model that adds the classification information of the problem to the probability of generation of the keyword in the document to smooth the probability of generation of the language model on the classification, which optimized the correlation calculation method of question retrieval. Zhang [7] proposed a topic translation model for question retrieval, which uses topic information as an implicit semantic constraint to adjust the correlation between words under the translation model, and thus effectively suppress translation noise.

## III. COMBINING Q&A PAIR QUALITY AND QUESTION RELEVANCE FEATURES ON COMMUNITY-BASED QUESTION RETRIEVAL

### A. Framework of the model

This paper uses the current state-of-the-art sorting learning model to train the quality and question relevance model features through fusion question and answer. The framework is shown in *Fig 2*:

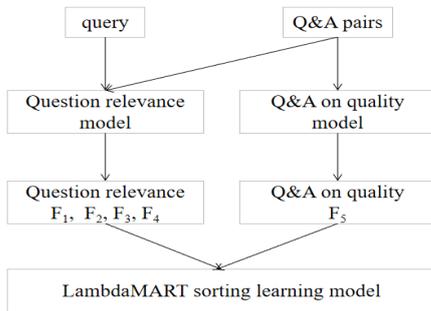

Fig. 2. Fusion Q&A on the quality of sorting learning model framework

Learning sorting is a supervised machine learning method that can easily fuse multiple features with fewer artificial parameters. From the current research methods, there are three strategies for learning sorting, namely pointwise, pairwise and listwise. The pointwise method converts the sorting problem into a multi-class classification or regression problem. The pairwise method converts the sorting problem into a document-to-classification problem. The listwise method learns the correct ordering of the entire candidate set under the query. The model used in this paper is LambdaMART. The advantage is that it can convert unrecognizable information retrieval evaluation indicators, such as NDCG, into functions that can be derived.

The LambdaMART[8] model can be split into Lambda and MART by name. MART is used as the bottom training model (Multiple Additive Regression Tree). Lambda is the gradient used by the MART solver process. Its physical meaning is the direction (up or down) and intensity of the next iteration of the document to be sorted. The specific algorithm process of LambdaMART is as follows: Firstly, the feature calculation method (described in detail below) is used to calculate the correlation feature scores of all question and answer pairs and query questions. Secondly, according to the manual labeling relevance, the initial state of the document pair corresponding to each question is reordered according to the labeling result to obtain the optimal sorting result. Thirdly, according to the position change of the optimal ranking result and the initial sequence, the corresponding deltaNDCG is calculated, and further, the Lambda between each query question and the corresponding question and answer pair is calculated. Finally, according to the obtained Lambda value, the regression tree is trained and solved.

Since the model can convert the unrecognizable IR evaluation index into a function that can be derived by transforming the loss function, the specific situation of this method is shown in *Fig 2*.

### B. Feature calculation

#### 1) Relevance feature

Zhang's proposed topic translation model T²LM [7] uses topic information to control translation noise in the translation model.

$$\begin{aligned} P_{t^2lm}(w/(q,a)) = & \mu_1 P_{ml}(w/q) \\ & + \mu_2 \sum_{t \in q}(P_{tr}(w/t)P_{ml}(t/q)) \\ & + \mu_3 \sum_{t \in q}\left(\sum_{i=1}^{K}(P_{to}(w|z_i)P_{to}(t|z_i))P_{ml}(t|q)\right) \quad (1) \\ & + \mu_3 \sum_{t \in q}\left(\sum_{i=1}^{K}(P_{to}(w|z_i)P_{to}(t|z_i))P_{ml}(t|q)\right) \\ & + \mu_4 P_{ml}(w|a) \end{aligned}$$

$P_{to}(w|z_i)$ is the distribution probability of the word $w$ under the topic $z_i$. $K$ is the number of topics in the topic model. $\mu_1$, $\mu_2$, $\mu_3$ and $\mu_4$ are the parameters used to control the impact of each part and $\mu_1 + \mu_2 + \mu_3 + \mu_4 = 1$.

As can be seen from the formula, they actually use $P_{to}(w|z_i) * P_{to}(t|z_i)$ to calculate the relevance of the word $w$ and the word $t$ under the topic $z_i$. Then obtain the total relevance of the word $w$ and the word $t$ under the topic model by summing the results under all the topics. Finally, the correlation under the theme model is combined with the relevance under the translation model to achieve the purpose of suppressing translation noise. The problem is that the weight of each topic is same in the process of summarizing the relevance results under all the topics, which makes it possible to determine the relevance of the vocabulary once the topic model is trained to complete the topic model changed. At the same time, since the inter-vocabulary relevance of the translation model is also unchanged after the training is completed, T²LM cannot respond differently to the lexical relevance of different user queries, and its relevance in terms of vocabulary is static.

Therefore, we have improved the model of Zhang. It first uses the topic model to analyze the topic distribution of the user query, then uses the topic information of the query to determine the specific semantics of the word in the retrieval process, and then guides the correlation calculation between the words in the topic model to get more accurate question relevance. The lengthy natural statement brings a lot of noise to the retrieval based on the word bag model while accurately conveying the user's search intention. The method of scoring the weights of the terms in the query is one of the effective means to suppress the adverse effects caused by the noise words. Therefore, we incorporate the weights of the query terms into it.

T²LM+ first obtains the weight $W_{query,w}$ by querying the word $w$ in the *query* through the term weighting model, and then integrates the weight $W_{query,w}$ into the correlation calculation. Specifically, for the word $w$ in the *query*, the correlation with the question and answer pair $(q,a)$ is calculated according to the following formula:

We split the correlation of the four parts of T²LM+ into four correlation features. They are the $F_1(query,q)$ query-question language model, $F_2(query,q)$ query-question translation model, $F_3(query,q)$ query-question topic model, and $F_4(query,a)$ query-answer language model. Specifically, the four correlation features are calculated according to formulas (4) to (7) respectively. Here for the smoothing parameter $\lambda$, we set $\lambda$ to $1/(|q|+1)$ or $1/(|a|+1)$.

*2) Q&A on quality feature*

In addition to question relevance, another recognized indicator of the impact of ranking results is the quality of questions and answers. In the case where the relevance of the candidate questions is close, a high-quality question and answer should be placed in the front position to enhance the user experience. In the research on community-based question retrieval, the quality of question-and-answer is often based not on the question-and-answer text itself, but on the use of other data in the Q&A community, such as the number of support and anti-objectives. We use user information, specifically the number of answers the user has adopted as the best answer, and design a user-authorized scoring model. The model is as follows:

$$S_u = \frac{Min(\sqrt{A_u},20)}{20} \quad (2)$$

$A_u$ is the best answer for user $u$. We set an upper limit for the user's authoritative score to eliminate the possible adverse effects of outliers.

$$P_{t^2lm+}((w,query)|(q,a)) = \mu_1 W_{query,w} P_{ml}(w|q)$$
$$+ \mu_2 \sum_{t \in q}(P_{tr}(w|t)P_{ml}(t|q))$$
$$+ \mu_3 \sum_{t \in q}\left(\sum_{i=1}^{K}(P(z_i|query)P_{to}(w|z_i)P_{to}(t|z_i))P_{ml}(t|q)\right)$$
$$+ \mu_4 W_{query,w} P_{ml}(w|a) \quad (3)$$

$$W_{query,w} = \frac{-\sum_{i=1}^{K} P(z_i|query)P(w|z_i)\ln P(w|z_i)}{-\sum_{t \in query}\sum_{i=1}^{K} P(z_i|query)P(t|z_i)\ln P(t|z_i)} \quad (4)$$

$$F_1(query,q) = \prod_{w \in query}\left((1-\lambda)W_{query,w}P_{ml}(w|q) + \lambda P_{ml}(w|C)\right) \quad (5)$$

$$F_2(query,q) = \prod_{w \in query}\left((1-\lambda)\sum_{t \in q}(P_{tr}(w|t)P_{ml}(t|q)) + \lambda P_{ml}(w|C)\right) \quad (6)$$

$$F_3(query,q) = \prod_{w \in query}\left((1-\lambda)\sum_{t \in q}\left(\sum_{i=1}^{K}(P(query|z_i)P_{to}(w|z_i)P_{to}(t|z_i))P_{ml}(t|q)\right) + \lambda P_{ml}(w|C)\right) \quad (7)$$

$$F_4(query,a) = \prod_{w \in query}\left((1-\lambda)W_{query,w}P_{ml}(w|a) + \lambda P_{ml}(w|C)\right) \quad (8)$$

Then, based on the relationship between the question-questioner and the answer-responder, the user's authoritative scoring model is transformed into a question-and-answer quality feature:

$$F_5((query,a)) \leftarrow (S_{qu}, S_{au}) \quad (9)$$

The user $qu$ is the proposer of the question $q$, and the user $au$ is the respondent of the answer $a$.

*C. Algorithm Description*

In summary, we propose a user-based Q&A quality assessment algorithm as shown in *Fig 3*., and a community-based question retrieval for the quality and question relevance features as shown in Fig.4. method. Algorithm 1 scores its authority based on the user's best answer number first, and then based on the assumption that the quality of the user's published information is positively related to its authority, the question and answer of the authoritative evaluation of the questioner and the respondent is used as a question and answer pair quality feature. *Fig 4* uses the sorting learning to combine the question relevance feature ($F_1 \sim F_4$) and the question and answer quality feature ($F_5$) to form the community-based question retrieval method for the fusion question and answer on the quality and question relevance features.

| Q&A based on user information for quality assessment algorithm |
|---|
| Input: Q&A pair $(q,a)$ |
| Output: Score of Q&A pair $(q,a)$ |
| 1) For $(q,a)$, get the questioner $q$ of question $u_q$ |
| 2) For $(q,a)$, get the Respondent $a$ of answer $u_a$ |
| 3) For the questioner $u_q$ and the respondent $u_a$, according to the author's authoritative evaluation model, the authoritative scores $S_{uq}$ and $S_{ua}$ are calculated respectively. |
| 4) Combining $S_{uq}$ and $S_{ua}$, the resulting vector $(S_{uq}, S_{ua})$ is the score of the $F_5((q,a))$ |

Fig. 3. Q&A based on user information for quality assessment algorithm

**Community-based question retrieval method based on fusion question and answer on quality and question relevance**

Input: Candidate Q&A pairs set $C$ and *query*
Output: Q&A pairs set sort result S of set q
1) For $(q,a)$ in $C$, Calculating its quality feature $F_5((q,a))$
2) For $(q,a)$ in $C$, Using the question relevance model to obtain the relevance features $F(query,(q,a))$ of the *query*
3) Enter the quality score $F_5((q,a))$ and the correlation feature $F(query,(q,a))$ of all Q&A pairs in $C$ into the trained LambdaMART model to get the sort result $Rank_{query,C}$.

Fig. 4. Community-based question retrieval method based on fusion question and answer on quality and question relevance

## IV. EXPERIMENTAL RESULTS AND ANALYSIS

### A. Experimental data

We used the data set from NDBC CUP 2016 as experimental data. There are 578608 questions and 1,729,263 answers in the data set. Since there is no correlation mark between the query and the candidate in the dataset, and there is currently no standard Chinese question search data set with annotations, Zhang uses manual markup to obtain experimental data, so we adopt the same method. We first obtain 1500 questions from the data set as a standby query by random sampling method, and use Lucene as a search tool to obtain the top 500 related problem pairs for each query as its candidate questions. Then according to the query content and the top20 question and answer pair, 990 available queries are manually selected. Here we select the available queries based on the content and length of the query and the top20 candidate questions that are related or similar to the query. Finally, 990 queries are randomly divided into training data sets and test data sets in a 1:1 ratio. The training set is mainly used to adjust the model parameters, and the test set is mainly used to test the effect of the model.

In order to enrich our data, we supplemented the number of marker candidates for queries with less than 20 annotation candidates to 20 by random sampling and noise from the ensemble. In the experiment, we set the number of trees in LambdaMART to 50, the number of leaves to 4, the learning rate to be 0.2, and the minimum number of instances of a single leaf to be 30.

Finally, we let the caller manually label the relevance of the query to its corresponding candidate question. Specifically, two labelers are invited first, each of which puts one of the following three labels for each candidate question according to the labeling specification: "good (2)", "general (1)", and "poor (0) )". If the two callers cannot agree on a candidate, the third caller is invited to make the final decision (he can only choose one). In order to reduce the workload of manual markup, we only mark the top10 candidate questions for the running results of each model. The other unmarked default tags are 0.

### B. Experimental comparison system

In order to verify the effect of our proposed learning ranking model, we also use the classic model of question retrieval and the current state-of-the-art model as a comparison system. As follows:

1) *Vector space model (VSM) [1].*
2) *BM25 model [2].*
3) *Unary language model (LM) [4].*
4) *Translation-based language model (TLM) [5].*
5) *Intent-based language model (IBLM) [6].*
6) *Language model based on topic translation model ($T^2LM$) [7].*
7) In addition to the traditional ML model, we propose a convolutional neural network method based on attention mechanism （TextCNN-Attention）. The method converts the sorting problem into a classification problem, with the question and the optional question as the network input, and the relevance as the category label. The specific structure is shown in Fig 5.

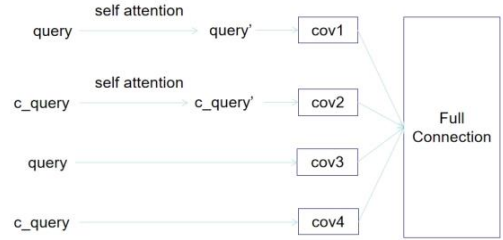

Fig. 5. TextCNN-Attention model framework

### C. Experimental results and analysis s

TABLE I. COMPARISON OF T2LM+ WITH EXISTING ADVANCED MODELS

|        | VSM    | BM25   | LM     | TLM    | IBLM   | $T^2LM$ | $T^2LM+$ |
|--------|--------|--------|--------|--------|--------|---------|----------|
| **MAP** | 0.3475 | 0.3506 | 0.3583 | 0.3746 | 0.3916 | 0.4361  | **0.4695** |
| **VSM** | N/A    | +0.31  | +1.08  | +2.71  | +4.41  | +8.86   | **+12.20** |
| **BM25** | N/A   | N/A    | +0.77  | +2.40  | +4.10  | +8.55   | **+11.89** |
| **LM**  | N/A    | N/A    | N/A    | +1.63  | +3.33  | +7.78   | **+11.12** |
| **TLM** | N/A    | N/A    | N/A    | N/A    | +1.70  | +6.15   | **+9.49** |
| **IBLM** | N/A   | N/A    | N/A    | N/A    | N/A    | +4.45   | **+7.79** |
| **$T^2LM$** | N/A | N/A   | N/A    | N/A    | N/A    | N/A     | **+3.34** |

TABLE II. EXPERIMENTAL RESULTS OF COMMUNITY-BASED QUESTION RETRIEVAL METHODS THAT COMBINE QUESTION AND ANSWER ON QUALITY AND QUESTION RELEVANCE

| MODEL           | MAP    |
|-----------------|--------|
| $T^2LM+$        | 0.4695 |
| $T^2LM+5$       | 0.4852 |
| TextCNN+Attention | 0.4992 |

As shown in Table 1, we first made a comparison of the $T^2LM$ model ($T^2LM+$) with the existing advanced model, and then we based on $T^2LM+$. We use $T^2LM+$ as four features ($T^2LM+5$) to conduct a fusion of Q&A on community-based question retrieval methods for quality and question relevance features. Finally, we compare the neural network model (TextCNN + Attention) with them.

From Table 2, we can see that the improved model ($T^2LM+$) is more effective than the topic-based translation query model ($T^2LM$), indicating that the improved model has a positive impact on question retrieval. At the same time, we can also see that the retrieval model based on learning sorting is stronger than T²LM+, which is mainly because the learning sorting LambdaMAR adopts the tree model, so it can learn the different combinations of features. We can also find that TextCNN has improved performance compared to traditional machine learning methods.

V. CONCLUSION

In view of the shortcomings of the traditional term weighting model, we propose a new topic-based model of term weighting model. It scores the importance of the term according to the amount of information, thereby highlighting the role of important words in the search and further optimizing the results of the question search.

The quality of questions and answers is considered to be an important indicator of the ranking results in addition to relevance. Just like PageRank is not based on web content, the quality of questions and answers can usually not be judged by question and answer text. In related research, researchers make full use of the diversity of data in the Q&A community, such as the number of support and anti-objectives, to assess the quality of questions and answers. In this paper, we propose a community-based question retrieval method that combines question and answer on quality and question relevance. We also proposed a deep learning question retrieval method. The experimental results validate the effects of our proposed model.

REFERENCES

[1] Salton G, Wong A, Yang C S. A vector s-pace model for automatic indexing[J]. Co-mmunications of the ACM, 1975, 18(11):613-620.

[2] Robertson S E, Walker S.Okapi/Keenbowa-tTREC-8.[J].TREC,1999,8: 151-162.

[3] Song F, Croft W B. A general language model for information retrieval[C]//Proceedi-ngs of the eighth international conference on Information and knowledge management.ACM,1999: 316-321.

[4] Ponte J M, Croft W B. A language modeli-ng approach to information retrieval[C]//Pr-oceedings of the 21st annual international A-CM SIGIR conference on Research and de-velopment in information retrieval.ACM,1998:275-281.

[5] Xue X, Jeon J, Croft W B. Retrieval models for question and answer archives[C]//Proceedings of the 31st annual international ACM SIGIR conference on Research and development in information retrieval.ACM,2008:475-482.

[6] Haocheng W. Research on Ranking Method in Community Question and Answer Search [D]. Anhui: University of Science and Technologyof China,2017.

[7] Weinan Z, Yu Z, Ting L. A topic translation model for community-based question retrieval [J]. Journal of Computer,2015, 38(2):313-321.

[8] Y. Wang, X. Lin, L. Wu, et al, Robust subspace clustering for multi-view data by exploiting correlation consensus. IEEE Transactions on Image Processing, 24(11):3939-3949, 2015.

[9] Y. Wang, L. Wu, X. Lin, J. Gao. Multiview Spectral Clustering via Structured Low-Rank Matrix Factorization. IEEE Transactions on Neural Networks and Learning Systems 29 (10), 4833-4843, 2018.

[10] Y. Wang, W. Zhang, L. Wu et al., Iterative Views Agreement: An Iterative Low-Rank based Structured Optimization Method to Multi-View Spectral Clustering. IJCAI 2016.

[11] L. Wu, Y. Wang. Beyond Low-Rank Representations: Orthogonal Clustering Basis Reconstruction with Optimized Graph Structure for Multi-view Spectral Clustering. Neural Networks, 103:1-8, 2018.

[12] Y. Wang, X. Lin, L. Wu, W. Zhang. Effective Multi-Query Expansions: Collaborative Deep Networks for Robust Landmark Retrieval. IEEE Transactions on Image Processing 26 (3), 1393-1404,

[13] L. Wu, Y. Wang, X. Li, J. Gao. Deep Attention-based Spatially Recursive Networks for Fine-Grained Visual Recognition. IEEE Transactions on Cybernetics 49 (5), 1791-1802, 2019.

[14] L. Wu, Y. Wang, L. Shao. Cycle-Consistent Deep Generative Hashing for Cross-Modal Retrieval. IEEE Transactions on Image Processing 28 (4), 1602-1612, 2019.

[15] Y. Wang, X. Lin, L. Wu et al., LBMCH: Learning Bridging Mapping for Cross-modal Hashing. ACM SIGIR 2015.

[16] huagong_adu. Learning To Rank. https://blog.csdn.net/huagong_adu/article/details/40710305.

[17] Brown P F, Pietra V J D, Pietra S A D, et al. The mathematics of statistical machine translation: parameter estimation[J].Computational Linguistics, 1993,19(2):263-311.

[18] Ponte J M, Croft W B. A language modeling approach to information retrieval[C]//Proceedings of the 21st annual international ACM SIGIR conference on Research and development in information retrieval.ACM, 1998:275-281.

[19] Xue X, Jeon J, Croft W B. Retrieval models for question and answer archives[C]//Proceedings of the 31st annual international ACM SIGIR conference on Research and development in information retrieval.ACM,2008:475-482.

[20] Lee J T, Kim S B, Song Y I, et al. Bridging lexical gaps between queries and questions on large online Q&A collections with compact translation models[C]//Proceedings of the Conference on Empirical Methods in Natural Language Processing. Association for Computational Linguistics,2008:410-418.

[21] Bernhard D, Gurevych I. Combining lexical semantic resources with question & answer archives for translation-based answer finding[C]//Proceedings of the Joint Conference of the 47th Annual Meeting of the ACL and the 4th International Joint Conference on Natural Language Processing of the AFNLP: Volume 2-Volume 2. Association for Computational Linguistics,2009:728-736.

[22] Gao J, Nie J Y, Wu G, et al. Dependence language model for information retrieval[C]. Proceedings of the 27th annual international ACM SIGIR conference on Research and development in information retrieval.2004:170–177.

[23] Cao X, Cong G, Cui B, et al. Approaches to exploring category information for question retrieval in community question-answer archives[J]. ACM Transactions on Information Systems,2012,30(2):7.